\documentclass[letterpaper]{article} % DO NOT CHANGE THIS
\usepackage{aaai2026}  % DO NOT CHANGE THIS
\usepackage{times}  % DO NOT CHANGE THIS
\usepackage{helvet}  % DO NOT CHANGE THIS
\usepackage{courier}  % DO NOT CHANGE THIS
\usepackage[hyphens]{url}  % DO NOT CHANGE THIS
\usepackage{graphicx} % DO NOT CHANGE THIS
\usepackage{natbib}  % DO NOT CHANGE THIS AND DO NOT ADD ANY OPTIONS TO IT
\usepackage{caption} % DO NOT CHANGE THIS AND DO NOT ADD ANY OPTIONS TO IT
\usepackage{algorithm}
\usepackage{algorithmic}
\usepackage{amsmath,amsfonts}
\usepackage{multirow}
\usepackage{makecell}
\usepackage{subcaption}
\usepackage[font={small}]{caption}
\def\eqref#1{equation~\ref{#1}}
\usepackage{amsmath}
\usepackage{amssymb}

\usepackage{amsthm}

\usepackage{tikz}
\usetikzlibrary{arrows}

\usepackage{booktabs}

\usepackage{newfloat}
\usepackage{listings}
\DeclareCaptionStyle{ruled}{labelfont=normalfont,labelsep=colon,strut=off} % DO NOT CHANGE THIS
\floatstyle{ruled}
\newfloat{listing}{tb}{lst}{}
\floatname{listing}{Listing}
\newcommand{\piechart}[5]{%
  \begin{tikzpicture}
    \pgfmathsetmacro{\grnEnd}{90 - #2*3.6}
    \pgfmathsetmacro{\yelEnd}{\grnEnd - #3*3.6}
    \pgfmathsetmacro{\redEnd}{\yelEnd - #4*3.6}
    \fill[green!55!black!80] (0,0) -- ({#1*cos(90)},{#1*sin(90)})
      arc (90:\grnEnd:#1) -- cycle;
    \fill[yellow!75!orange] (0,0) -- ({#1*cos(\grnEnd)},{#1*sin(\grnEnd)})
      arc (\grnEnd:\yelEnd:#1) -- cycle;
    \fill[red!65!black!90] (0,0) -- ({#1*cos(\yelEnd)},{#1*sin(\yelEnd)})
      arc (\yelEnd:\redEnd:#1) -- cycle;
    \fill[gray!35] (0,0) -- ({#1*cos(\redEnd)},{#1*sin(\redEnd)})
      arc (\redEnd:{90-360}:#1) -- cycle;
    \draw[white, line width=0.6pt] (0,0) -- ({#1*cos(90)},{#1*sin(90)});
    \draw[white, line width=0.6pt] (0,0) -- ({#1*cos(\grnEnd)},{#1*sin(\grnEnd)});
    \draw[white, line width=0.6pt] (0,0) -- ({#1*cos(\yelEnd)},{#1*sin(\yelEnd)});
    \draw[white, line width=0.6pt] (0,0) -- ({#1*cos(\redEnd)},{#1*sin(\redEnd)});
    \fill[white] (0,0) circle (#1*0.4);
    \draw[gray!40, line width=0.3pt] (0,0) circle (#1);
    \draw[gray!40, line width=0.3pt] (0,0) circle (#1*0.4);
  \end{tikzpicture}%
}

\usepackage{enumitem}
\usepackage{fontawesome5}
\usepackage{xcolor}
\definecolor{chev1}{RGB}{42,  75,  110}
\definecolor{chev2}{RGB}{30,  140, 140}
\definecolor{chev3}{RGB}{60,  160, 100}
\definecolor{chev4}{RGB}{220, 160,  40}
\definecolor{chev5}{RGB}{200,  75,  55}

\newcommand{\mychevron}[3]{%
  \begin{scope}[shift={(#1,0)}]
    \ifnum#3=1
      \filldraw[fill=#2, draw=white, line width=1.5pt]
        (0,0) -- (2.2,0) -- (3.0,1.0) -- (2.2,2.0) -- (0,2.0) -- cycle;
    \else
      \filldraw[fill=#2, draw=white, line width=1.5pt]
        (0,0) -- (2.2,0) -- (3.0,1.0) -- (2.2,2.0) -- (0,2.0)
        -- (0.8,1.0) -- cycle;
    \fi
  \end{scope}
}

\title{Are Algorithm Registers Transparent? Perspectives from Germany}

\author {
    Iman Peljto\textsuperscript{\rm 1,\rm 2},
    Xenia Heilmann\textsuperscript{\rm 2},
    Mattia Cerrato\textsuperscript{\rm 2}
}
\affiliations {
    \textsuperscript{\rm 1}TU Darmstadt\\
Germany\\
 
    \textsuperscript{\rm 2} Johannes Gutenberg University Mainz\\  Germany\\
    iman.peljto@stud.tu-darmstadt.de, xenia.heilmann@uni-mainz.de, mcerrato@uni-mainz.de
}

\begin{document}

\maketitle

\begin{abstract}
%Old
%Algorithm registers are public-facing databases that display basic information about algorithms employed in public administration. There are several registers available today, especially relating to usage of AI in government: examples are the Dutch register and the Eurocities project. The EU AI Act itself mandates a database for ``high-risk'' applications. In this paper, we employ a recent proposal by \citet{algorithmwatch} detailing the requirements and objectives for a German algorithmic and AI register. Based on this proposal, we develop checklists of features, both at the technical and governance level, that should be implemented in such a register. Then, we employ this framework to evaluate two existing transparency initiatives in Germany, MaKI and Lernende Systeme. We find that several adaptations are likely needed for these registers to serve as an useful transparency instrument. 

%New Proposal
Algorithm registers are public-facing databases that display basic information about algorithms employed in public administration.  While several such registers exist across Europe and globally, their capacity to deliver meaningful transparency remains contested. In Germany, the landscape is notably fragmented: no federal-level register exists, yet at least five state- and federal-level initiatives publish information about AI systems with varying scopes and objectives. A recent conceptual proposal by~\citet{algorithmwatch} outlines technical and governance requirements for a national AI transparency register in Germany. We repurpose this proposal as an audit instrument, extracting structured checklists from the transparency goals and subgoals formulated by~\citet{algorithmwatch}. The resulting checklists, translated from German into English, are made publicly available to support practitioners auditing existing registers or designing new ones. We apply this framework to conduct an external audit of the two main existing German transparency initiatives, \textit{Marktplatz der KI-Möglichkeiten} and \textit{Plattform Lernende Systeme}, evaluating the extent to which they fulfill the proposed goals. Our audit reveals that several adaptations are likely needed for these registers to serve as a useful transparency instrument. We further propose a visualization of register transparency levels and derive concrete action items for improving existing German platforms. 
\end{abstract}
\begin{links}
    %\link{Code}{https://aaai.org/example/code}
    %\link{Datasets}{https://aaai.org/example/datasets}
    \link{Extended version}{https://arxiv.org/abs/2606.02347}
\end{links}

\section{Introduction}

Algorithm registers are publicly accessible databases documenting algorithmic systems. \citet[p.~16]{murad_2021} defines a register as ``a log of algorithmic decision making systems used by a public authority that have some level of direct impact on its citizens.'' Of particular interest are registers maintained by government bodies:  Available registers and standards include the Dutch AI register~\cite{cath_dutch_2022}, the UK~\cite{open_government_partnership_promoting_2025}, and several others in the EU~\cite{soizic_penicaud_making_2025} and globally~\cite{global_partnership_on_artificial_intelligence_algorithmic_2024}. These national transparency initiatives are increasingly relevant in light of the supra-national EU database for high-
risk AI systems listed in Annex III of the EU AI Act and
mandated under Article 71 (European Parliament and Council 2024). Interoperability between national transparency
registers and the future EU database could substantially facilitate compliance and reporting obligations across EU member states. 
In the European setting, the German situation is notable for its fragmentation. While the Bundestag's Digital Committee has expressed interest in a federal-level register and has conducted a public hearing that touched upon its foundation~\cite{david_roth-isigkeit_stellungnahme_2024}, we are not aware of a concrete plan towards the establishment of a new register. Rather, there exist at least five state- and federal-level initiatives that deal with the publication of information about AI systems, each with different scopes and objectives (see Section~\ref{sec:fragmented}).

This fragmented landscape has recently received a notable conceptual contribution from~\citet{algorithmwatch}, who presents a technical and organizational concept for a national AI transparency register in Germany. The project was funded by the VolkswagenStiftung and the work is published by AlgorithmWatch. Lorenz's report starts from a description of the current limitations in federal-level efforts~\citep[pp.~4--6]{algorithmwatch} and proposes an alternative at both the technical and governance levels. In this paper, we repurpose Lorenz's proposal as an audit instrument: rather than using it to build a new register, we extract from it a checklist of goals and subgoals and apply it to two existing German platforms, \textit{Marktplatz der KI-Möglichkeiten} (MaKI) and \textit{Plattform Lernende Systeme} (PLS) (described in Section~\ref{sec:platforms}). Our purpose is to evaluate the \emph{transparency} of existing \emph{transparency initiatives} in Germany. Our main finding is that several gaps exist across multiple dimensions, particularly at the level of training data documentation and risk assessments.
To summarize, our main contributions are as follows:
\begin{enumerate}
  \item We review the European algorithm register literature and organize it along three dimensions of concern (scope, information depth, and effectiveness) that inform our subsequent audit framework.
  \item We develop an audit approach structured around the transparency goals suggested by~\citet{algorithmwatch} and translate the resulting checklists from German into English (see Appendix~B%\ref{app:checklists}
  \footnote{See the extended version: \url{https://arxiv.org/abs/2606.02347.}}), facilitating reuse by practitioners aiming to audit existing registers or develop new ones.
  \item We apply this approach to conduct an external audit of MaKI and PLS, analyzing the extent to which the primary and secondary goals of~\citet[Table~39]{algorithmwatch} are fulfilled.
  \item We propose a visualization of register transparency levels and identify concrete action items towards greater fulfillment of transparency goals.
\end{enumerate}

\section{Algorithm Registers in Europe: Scope, Standards, and Limitations}\label{sec:related-work}

Among the transparency instruments available to governments, algorithm registers occupy a distinctive position. \citet{soizic_penicaud_making_2025} classify them as \emph{systemic, supply-driven} instruments: proactive disclosures that aim to make an entire portfolio of systems visible, shifting the burden of initiative from the public to the institution.

Amsterdam and Helsinki became the first cities to publish such registers in September 2020~\cite{floridi_artificial_2020, haataja_public_2020}. Adoption has since accelerated: the Eurocities network developed a shared Algorithmic Transparency Standard involving nine cities~\cite{noauthor_algorithm_nodate}; the UK launched its Algorithmic Transparency Recording Standard (ATRS) in 2021 and made it mandatory for all central government departments in 2024~\cite{open_government_partnership_promoting_2025, leslie_securing_2024}; and the Netherlands launched a national register (\texttt{algoritmes.overheid.nl}) in December 2022. By 2024, the Global Partnership on AI had mapped 69 active repositories worldwide~\cite{global_partnership_on_artificial_intelligence_algorithmic_2024}, while \citet{soizic_penicaud_making_2025} counted 34 registers in the EU alone. At the EU level, the AI Act~\cite{AIA2024} mandates a public database of high-risk AI systems (Article~71), though systems used in law enforcement, migration, asylum, and border control are registered in a non-public section (Article~49(4)).

Despite this rapid proliferation, the literature reveals recurring concerns about the design and effectiveness of these instruments. We organize these concerns along three dimensions: the \emph{scope problem} (what gets registered), the \emph{information problem} (what is disclosed about each registered system), and the \emph{effectiveness problem} (whether registers achieve their stated goals in practice). %These three dimensions structure both our review and, in later sections, the checklist we derive from Lorenz's proposal.

\subsection{The Scope Problem: What Gets Registered?}

A foundational design question for any register is what counts as an ``algorithm'' or ``AI system'' that should be registered. The AI Act's definition (Article~3(1)) centers on machine-based systems that operate with varying levels of autonomy and that infer how to generate outputs from their inputs. The European Commission published non-binding guidelines in February 2025 attempting to clarify the boundary, citing linear and logistic regression as explicit examples of non-AI systems~\citep[Paragraph~46]{european_commission_guidelines_2025} while including symbolic methods such as expert systems~\citep[Paragraph~39]{european_commission_guidelines_2025}. The guidelines ultimately acknowledge that ``no automatic determination or exhaustive lists of systems that either fall within or outside the definition of an AI system are possible'' (\emph{ibid.}, Paragraph~62). \citet{soizic_penicaud_making_2025} flag this ambiguity as a direct concern for registers: if the definition excludes rule-based automation, then systems that may carry substantial risks, such as deterministic scoring or eligibility algorithms, would fall outside the scope of both the EU database and any national register that adopts the AI Act definition. \citet[pp.~6--7, 18--19]{murad_2021} frames the issue more broadly as a question about the appropriate ``unit of disclosure,'' arguing that transparency should apply to the algorithmic decision-making system as a whole rather than to the algorithm in a narrow technical sense.

Beyond the definition of what \emph{qualifies} for registration, there is the question of what \emph{actually gets registered} in practice. \citet[Sec.~1--2]{cath_dutch_2022} show how the Amsterdam and Helsinki registers were narrowly scoped to cover ``a set of largely uncontentious bureaucratic municipal uses of AI'' while excluding sensitive domains such as welfare fraud detection and predictive policing. They argue that these selective inclusions and exclusions are themselves political acts. \citet[p.~418]{nieuwenhuizen_algorithm_2024} generalizes this observation, noting that the unit of disclosure is ``inherently a political choice, impacting the register's comprehensiveness and reflecting the organization's values and priorities.'' Her empirical study found that more than half of Dutch municipalities had registered nothing at all. A further scope question is whether to include systems still in development or already decommissioned: the Dutch standard permits but does not require pre-deployment registration; the Eurocities standard provides a seven-stage lifecycle from \emph{planned} through \emph{retired}; and \citet{soizic_penicaud_making_2025} and the Ada Lovelace Institute et al.~\cite{ada_lovelace_institute_algorithmic_2021} advocate for pre-deployment registration as a condition for early citizen participation.

\subsection{The Information Problem: What Is Disclosed?}

Even when a system is registered, the question remains: what information is captured, and in what form? The two most developed European standards illustrate the range of approaches. Other national registers, such as Scotland's AI Register\footnote{\url{https://scottishairegister.com}} and Norway's public-sector AI overview,\footnote{\url{https://data.norge.no/kunstig-intelligens}} are not analyzed separately here: Scotland's register is built on the same technical platform used by the Amsterdam and Helsinki registers~\cite{haataja_public_2020} and does not define an independent schema; Norway's overview is a catalog of projects within a broader national data portal, without published field definitions that could be compared against the Dutch or Eurocities schemas.

The \textbf{Dutch National Publication Standard}~(v1.0.10) \cite{minbzk_publicatiestandaard_2024} organizes information into four sections: General Information, Responsible Use, Functioning, and Metadata. It contains no dedicated training data field; the closest approximation is the free-text ``Gegevens'' (Data) field, which does not distinguish training data from operational input and imposes no structure on data provenance, collection period, or bias assessment. The ``Technische werking'' (Technical functioning) field must indicate whether the system is self-learning but provides no further sub-fields. The stated objective is to protect ``people, their fundamental rights, and public values'' and to ``enable citizens to critically monitor and question the government.''\footnote{\url{https://www.nldigitalgovernment.nl/overview/algorithms/}}

The \textbf{Eurocities Algorithmic Transparency Standard}~(v0.3.1) \cite{noauthor_algorithm_nodate} is more ambitious. It introduces \texttt{training\_data} as a named, required field separate from \texttt{source\_data} (operational input). An optional \texttt{data\_bias} field explicitly asks whether data contain biases and how these are compensated for. The Oversight category mandates documentation of performance monitoring, human intervention procedures, type of use (descriptive, diagnostic, predictive, or prescriptive), and objection mechanisms. However, all fields remain free-text prose without structured sub-fields, and the ``required'' designation carries no enforcement since participation is voluntary. \citet{leslie_securing_2024}, comparing transparency requirements across five jurisdictions, find that no country has yet established legally enforceable, comprehensive disclosure standards for public-sector AI, reinforcing the concern that even the most developed register standards remain voluntary instruments.

A notable gap across both standards, and across the German proposal discussed in Section~\ref{sec:lorenz-report}, is the absence of any field for attaching or referencing model cards~\cite{mitchell_model_2019} or datasheets for datasets~\cite{gebru_datasheets_2021}. Including such a mechanism would enable layered disclosure, where a general audience reads the register entry while a more technical audience can access structured, model-level documentation.

\subsection{The Effectiveness Problem: Do Registers Work?}

The most fundamental criticism of algorithm registers is that their existence may create a misleading impression of accountability. \citet[Sec.~4]{cath_dutch_2022} warn against ``romanticizing the light touch ex-post transparency measures'' provided by registers, arguing that academic enthusiasm about their mere existence can produce what they term ``ethics theater'': visible but substantively hollow measures that create a false sense of oversight. \citet{busuioc_reclaiming_2023} raise a related concern in their analysis of the EU AI Act: substituting transparency with mediated explanations, where the provider controls what is made visible, opens disclosure to manipulation. In a register context, this applies directly, since the registering body is typically also the deployer and no independent verification is required.

A complementary line of criticism asks: \emph{transparent to whom}? \citet{kemper_transparent_2019} argue that transparency only becomes accountability when a \emph{critical audience} exists that is capable of interpreting and acting on the disclosed information. \citet{bell_think_2023} build on this reasoning, proposing that disclosure mechanisms should be designed based on what different audiences need. \citet{nieuwenhuizen_algorithm_2024} provides the most direct empirical evidence: oversight authorities and civil society watchdogs in the Netherlands reported that the information disclosed was of limited usefulness, and only 5\% of entries included a fundamental rights assessment. She concludes that transparency intermediaries should be the targeted audience if external accountability is the goal.

Yet Nieuwenhuizen's findings are not uniformly negative. She also documents a \emph{disciplining effect}: the process of registration compels civil servants to reflect on whether a system is justified, what its legal basis is, and how risks are managed. Her characterization of registers as a ``meaningful box-ticking exercise'' captures the duality: the exercise creates organizational value even when its public-facing output is thin. Similar findings from six case studies at Dutch public organizations are reported by~\citet{van_vliet_defining_2024}, who propose a reference process for register implementation. \citet{soizic_penicaud_making_2025} distill from their survey of 34 EU registers a set of design recommendations, including centralized mandatory registration, collaboration with civil society in defining categories, coverage of in-development and decommissioned systems, and a requirement that registering bodies document and justify the absence of information rather than leaving fields empty.

The recurring finding across this literature is that registers function better as internal governance tools, prompting reflection and organizational learning, than as genuine public accountability mechanisms enabling external scrutiny. This gap between aspiration and practice motivates the present paper: rather than proposing yet another standard, we take a recent comprehensive proposal for a German national register~\cite{algorithmwatch} and derive from it a checklist of goals, which we then use to evaluate the existing German initiatives.

\section{Proposal for a German National AI Register}\label{sec:lorenz-report}

\citet{algorithmwatch} presents a detailed concept for the institutional, organizational, and technical implementation of a German national AI transparency register (KI-Transparenzregister). It is, to our knowledge, the most comprehensive proposal for a German register, covering not only the information schema but also governance structures, technical architecture, and stakeholder engagement. For this reason, we use it as the basis for our evaluation framework. We summarize its main features below before describing, in Section~\ref{sec:methodology}, how we transform its goals and subgoals into a checklist for evaluating existing registers.

The report is written in German. The German institutional names and technical terms used below follow Lorenz's original text; English translations are ours unless otherwise noted.

\subsection{Scope and Design Choices}
\label{sec:fragmented}

\citet{algorithmwatch} begins by surveying the fragmented German landscape of AI-related platforms, identifying five federal- and state-level initiatives (described in Section~\ref{sec:platforms}): MaKI, PLS, the KI-Observatorium, Mission KI, and KI.NRW. These differ substantially in scope, audience, and standards, and none functions as a comprehensive public transparency register. Furthermore, since December 2025, Mission KI is no longer operating. While Lorenz does not offer a full evaluation of MaKI, she comments on the lack of obligatory reporting and the lack of coordination between platforms~\citep[pp.~4--5]{algorithmwatch}: the same AI applications sometimes appear on multiple platforms under different names or with divergent descriptions.

From this assessment, Lorenz defines five target groups (citizens, public authorities, supervisory bodies, civil society and academia, and private-sector organizations) and five primary goals: safeguarding fundamental rights, strengthening public trust, enabling dialogue and review, centralizing AI system information, and ensuring interoperability with the EU high-risk AI database. Three secondary goals concern knowledge sharing, cross-sector cooperation, and minimizing administrative burden. The interoperability goal is notable: Lorenz's proposal is designed as a national complement to the EU database mandated by Article~71 of the AI Act, intended to go beyond it in both coverage (including systems below the high-risk threshold) and information richness.

Several further design choices are worth highlighting. First, Lorenz aligns the definition of AI systems with the EU AI Act, arguing that this facilitates compliance, comparability, and legal certainty. Second, she restricts coverage to systems that are or have been deployed (``produktiv''), explicitly excluding systems in planning or development. Her rationale is to maximize public relevance and avoid the problem she observes in MaKI, where only a quarter of listed systems were operational~\footnote{As of May 2026, this still holds. Currently, only two-thirds of MaKI's entries are operational.}. This stands in contrast with the Dutch register (which permits pre-deployment registration), the Eurocities standard (which provides a full seven-stage lifecycle), and the recommendations of both \citet{soizic_penicaud_making_2025} and the Ada Lovelace Institute et al.~\cite{ada_lovelace_institute_algorithmic_2021}, who advocate for pre-deployment registration as a condition for early citizen participation. Third, Lorenz explicitly declares that the register should not itself serve as a mechanism for reviewing ethical, legal, or data protection compliance; it should document such reviews, not conduct them. Her own schema, however, includes nine fields dedicated to risk classification, risk assessment results, and impact assessment outcomes, creating a tension between documenting compliance outputs and disclaiming responsibility for their quality.

\subsection{Information Schema}

The proposed schema contains 30 fields, divided into externally visible and internal fields, organized across four categories: information about the responsible institution (Informationen zur verantwortlichen Institution), information about the AI system (Informationen zum KI-System), information on criticality (Informationen zur Kritikalit\"{a}t des KI-Systems), and metadata about the submitting person (Angaben zur \"{u}bermittelnden Person). We provide the translated field names and descriptions in the Appendix, Table~3.%\ref{tab:informationfields}.

The treatment of training data is minimal. Field~16 calls for a ``basic and concise description of the information used and the operating logic'' (transl. from ``Grundlegende und knappe Beschreibung der verwendeten Informationen (Daten, Eing\"{a}nge) und der Betriebslogik''), citing Articles~51 and~53 of the AI Act. This field does not distinguish between training data and operational input, and provides no structured sub-fields for dataset provenance, size, collection period, or bias assessment. In this respect, the proposal is comparable to the Dutch national standard and less detailed than the Eurocities standard, which mandates separate \texttt{training\_data} and \texttt{source\_data} fields and offers an optional \texttt{data\_bias} field. Field~16 does include a provision for general-purpose AI systems, where the free-text description may be replaced by an upload of a training data summary template (``Zusammenfassung der Trainingsdaten'').

The criticality section is more developed, comprising nine fields covering risk classification under the AI Act, risk assessment, identified risks, mitigation measures, data protection impact assessment, and fundamental rights impact assessment. Across these fields, the same logic is applied: one field documents whether and how an assessment was conducted, while the corresponding field captures its results. Lorenz recommends following established risk assessment frameworks (ISO~31000:2018, ISO/IEC~23894:2023, NIST AI RMF, or IEEE EAD) and connecting register entries to the risk identification and evaluation steps prescribed therein.

\subsection{Architecture, Governance, and Outreach}

A distinguishing feature of the proposal is its depth on technical and institutional matters. On the technical side, Lorenz describes distinct user interfaces for data entry, analysis, reporting, and feedback, and devotes attention to what she terms \emph{bitemporale Historisierung} (bitemporal historization), the time-stamping and versioning of all register data. The reporting function envisions individual AI system profiles (KI-Steckbriefe), geographic maps of AI deployment (KI-Landkarten), aggregated statistics, and an API supporting automated analyses and integration with other databases. Any user of the register may submit feedback, which is connected to an evaluation of the register's own impact.

On the governance side, Lorenz proposes a tripartite division of responsibility: professional (setting guidelines and monitoring requirements), organizational (coordinating activities and stakeholder relations), and technical (infrastructure operation, potentially outsourced). She notes ongoing discussions about designating the Bundesnetzagentur\footnote{The Bundesnetzagentur f\"{u}r Elektrizit\"{a}t, Gas, Telekommunikation, Post und Eisenbahnen is Germany's multi-sector infrastructure regulator. It does not offer a translation of its name to English (see \url{https://www.bundesnetzagentur.de/EN/}).} as the coordinating body for national AI supervision, with a new dedicated digital agency as an alternative.

Finally, Lorenz integrates information campaigns, workshops, best practice templates, and citizen forums (B\"{u}rgerforen) into the register concept itself, treating outreach and capacity building as a component of the register rather than an afterthought. This echoes \citet{soizic_penicaud_making_2025}'s recommendation for active dissemination and responds to the empirical finding that registers require informed audiences to produce accountability effects~\cite{kemper_transparent_2019, nieuwenhuizen_algorithm_2024}.

\subsection{From Proposal to Evaluation Framework}

Lorenz's report is a normative proposal rather than an empirical study. Its value for our purposes lies not in whether its recommendations are optimal, but in the fact that it articulates, in concrete and operational terms, what a comprehensive German AI register should contain and how it should function. By taking the proposal at face value and extracting from it a set of (sub)goals, we obtain a checklist that can be applied to existing German initiatives to assess the extent to which they fulfill their stated transparency objectives. This approach has limitations: the checklist inherits any blind spots in Lorenz's proposal, and it evaluates registers against their purported goals rather than against an ideal standard. We accept these limitations deliberately. We aim to provide a structured, reproducible method for evaluating the registers that exist.

\section{Existing AI Platforms in Germany}\label{sec:platforms}

Having summarized Lorenz's proposal and her assessment of the current landscape, we now describe the two platforms that form the subject of our audit.

\subsection{Marktplatz der KI-Möglichkeiten}
The “Marktplatz der KI-Möglichkeiten” (MaKI)\footnote{\url{https://maki.beki.bund.de}} describes itself as a central AI transparency register for the federal public administration. According to its official mission, the platform aims to provide a comprehensive overview of AI use across the German public administration, facilitate exchange on AI systems and applications, and promote the reuse of existing solutions \cite{maki}. It provides detailed and accessible information on individual AI applications, referred to as \textit{KI-Steckbriefe} (AI fact sheets). 
Launched in January 2025, MaKI was conceived as a central coordination and networking tool for AI use within the federal administration. 
At its introduction, the platform listed only forty-six productive AI applications, with participation remaining voluntary. 
As of May 2026, 115 active systems are described, with 205 more at various stages of planning and decommissioning.
The model of registration is voluntary: as such, our evaluation is that rather than promoting transparency to the public, MaKI primarily serves internal purposes, connecting administrative bodies dealing with AI systems and supporting internal knowledge exchange. 

\subsection{Plattform Lernende Systeme (Platform Learning Systems)}
The  ``Plattform Lernende Systeme'' (PLS)\footnote{\url{https://www.plattform-lernende-systeme.de}} is an initiative funded by the Federal Ministry of Education and Research (BMBF) of Germany. It serves primarily as an exchange and dialogue platform for AI research, innovation, and application development in Germany. Its focus is on the scientific and industrial advancement of AI, not on governmental or administrative use, which distinguishes it clearly from initiatives like MaKI. PLS collects and presents examples of AI systems developed or deployed in Germany, emphasizing innovation transfer, best practices, and cross-sector learning. Beyond individual case studies, the platform also provides a searchable and filterable overview of registered AI systems, which can be explored either through an interactive \textit{KI-Landkarte} (AI map) or in a structured list view. The \textit{KI-Landkarte} visualizes the geographical distribution of AI systems across Germany, helping to identify regional concentrations, innovation clusters, and areas of potential policy or research interest. At the time of data extraction, the platform contained 1,305 registered AI projects in Germany, together with structured metadata on their application domain, technology field, industry sector, funding context, and developer information. 
\subsection{Other Platforms}
Beyond MaKI and PLS, several smaller platforms serve various purposes, namely the KI‑Observatorium, Mission‑KI, and KI.NRW. According to \citet{algorithmwatch}, KI‑Observatorium and Mission‑KI cannot be regarded as initiatives on the road to such a register. Nevertheless, both platforms provide useful products, information, and statistics on AI‑related topics for the general public. For example, Mission‑KI has developed a quality standard for low‑risk AI, but has stopped operation in December 2025 after a change in the German government. KI‑Observatorium has published articles on AI trends and terminology. KI.NRW offers an AI map of applications in North Rhine‑Westphalia; however, because it covers only a single federal state, it does not provide information at the national level. Moreover, as the platform is essentially a map, populating the tables required for this analysis yields few entries. Consequently, MaKI and PLS remain the two primary platforms examined in this paper.

\section{Methodology}

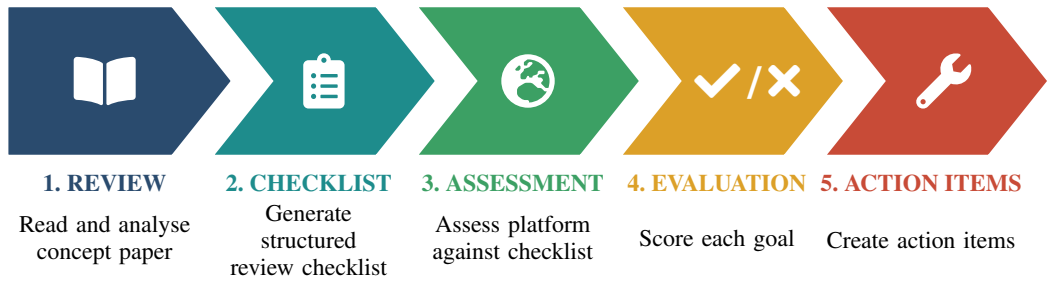
\begin{figure*}[t]
\centering
\begin{tikzpicture}

  \mychevron{0.0} {chev1}{1}
  \mychevron{2.7} {chev2}{0}
  \mychevron{5.4} {chev3}{0}
  \mychevron{8.1} {chev4}{0}
  \mychevron{10.8}{chev5}{0}

  \node[white] at (1.3, 1.0) {\Huge\faBookOpen};
  \node[white] at (4.2, 1.0) {\Huge\faClipboardList};
  \node[white] at (6.9, 1.0) {\Huge\faGlobeEurope};
  \node[white] at (9.8, 1.0) {%
    \huge\faCheck\,\textbf{/}\,\faTimes};
  \node[white] at (12.4,1.0) {\Huge\faWrench};

  % labels
  \node[font=\bfseries\small, text=chev1] at (1.3, -0.35) {1. REVIEW};
  \node[font=\bfseries\small, text=chev2] at (4.0, -0.35) {2. CHECKLIST};
  \node[font=\bfseries\small, text=chev3] at (6.7, -0.35) {3. ASSESSMENT};
  \node[font=\bfseries\small, text=chev4] at (9.4, -0.35) {4. EVALUATION};
  \node[font=\bfseries\small, text=chev5] at (12.1,-0.35) {5. ACTION ITEMS};

  % descriptions
  \node[text width=2.4cm, align=center, font=\small]
    at (1.3, -1.1) {Read and analyse concept paper};
  \node[text width=2.4cm, align=center, font=\small]
    at (4.0, -1.1) {Generate structured review checklist};
  \node[text width=2.4cm, align=center, font=\small]
    at (6.7, -1.1) {Assess platform against checklist};
  \node[text width=2.4cm, align=center, font=\small]
    at (9.4, -1.1) {Score each goal};
  \node[text width=2.6cm, align=center, font=\small]
    at (12.1,-1.1) {Create action items};

\end{tikzpicture}
\caption{Methodology overview.}
\label{fig:methodology}
\end{figure*}
\label{sec:methodology}
Lorenz's concept paper was written as a development guide: it specifies what a future German register should contain and how it should be built. We repurpose it as an audit instrument, using its requirements not to build a register but to evaluate existing ones. This shift in perspective is deliberate. The development guide articulates what \emph{should} be present; an audit checks what \emph{is} present. The gap between the two reveals the distance that existing platforms must travel to meet the transparency goals of algorithm registers. 
This reorientation introduces a structural limitation: some of Lorenz's subgoals presuppose access to internal architecture, governance processes, or implementation decisions that are not visible from outside. Rather than discarding these items, we retain them and classify them as \textit{No Access}, so that the limits of what an external audit can assess become visible in the results. This visibility can itself support accountability, particularly in cases where restricted access may not be clearly justified by privacy, security, or safety concerns. We proceed in five steps (Figure~\ref{fig:methodology}).

\subsection{Step 1: Review}
As a first step, we review and translate the concept paper by~\citet{algorithmwatch} into English. During this process, we analyze each section individually to identify its core objectives and the normative framework it establishes for transparency registers. We also summarize the overarching requirements articulated throughout the paper. The resulting synthesis forms the basis of Section~\ref{sec:lorenz-report} and provides the conceptual foundation for all subsequent steps.

\subsection{Step 2: Audit Checklists}\label{sec:checklist}
Although the concept paper by~\citet{algorithmwatch} was originally conceived as development guidelines for a national transparency register, we repurpose its content for the context of auditing existing registers. This shift in perspective requires translation of prescriptive design criteria into audit items which can be evaluated empirically. For each section of the concept paper, we identify the overarching requirements and supporting subgoals needed to achieve the intended goals, then translate them into structured checklist items. This preserves the original structure of the paper while enabling systematic comparison across platforms. Where requirements are formulated abstractly, we operationalize them into concrete indicators to the extent that the source material permits. The resulting checklists are provided in English in the Appendix~B, %\ref{app:checklists},
Table~3 %\ref{tab:informationfields}
to~9 %\ref{tab:scope},
with each table corresponding to one chapter of the underlying concept paper. In addition, we extract the primary and secondary goals of a transparency register as a whole, as defined in \citet[Table~39]{algorithmwatch}, and render these in checklist format as well.

\subsection{Step 3: Platform Assessment}\label{sec:assesment}

We apply the checklists developed in Step 2 to the two German platforms under review: MaKI and PLS. For each checklist item, we compare the requirements defined by \citet{algorithmwatch} against the information publicly available on each platform, drawing exclusively on content accessible to any external evaluator without special authorization or technical access. It should be noted that a recurring obstacle throughout this assessment is restricted access to certain data. This is a structural consequence of our methodological reorientation: we employ the proposal by \citet{algorithmwatch} as an auditing instrument, whereas it was originally designed as a development guide. As a result, some requirements presume insider knowledge of system architecture, internal processes, or implementation decisions that MaKI and PLS do not disclose publicly. %This tension between the concept paper's original scope and its application as an audit framework shapes the boundaries of our assessment throughout.
Each checklist item is classified into one of four categories: \textit{Yes}, \textit{No}, \textit{Partially}, and \textit{No Access}. Together, these classifications yield a comprehensive picture of which requirements are satisfied, partially satisfied, or impossible to assess for the public. The full results are documented in the Appendix~B %\ref{app:checklists}
Table~3 %\ref{tab:informationfields} 
to~9.%\ref{tab:scope}.

\subsubsection{Yes}
A criterion is marked \textit{Yes} when the required information or feature is directly locatable and observable on the platform in a form that fully satisfies the stated requirement. This classification is applied only to criteria that can be evaluated on the basis of observable content alone, i.e., the presence of a system name, provider information, a description of the AI system's purpose, or publicly accessible explanatory material.

\subsubsection{No}
A criterion is marked \textit{No} when the required information or feature is absent from the platform, and where its presence would reasonably be expected to be publicly verifiable. This classification signals a documentable gap as the platform does not indicate that the requirement is addressed anywhere on its publicly accessible interface.
\subsubsection{Partially}
\textit{Partially} fulfilled denotes the classification applied when a criterion is observable on the platform and shows evidence of implementation, but fails to meet the full scope of the goal due to incomplete execution, missing components, or functionality that cannot be verified in its entirety. This category captures insufficient compliance: the platform has engaged with the requirement but has not fully discharged it.
For instance, MaKI includes dedicated fields for data protection impact assessments under GDPR Article 35, yet these fields remain unpopulated for all analyzed AI systems, and no public justification is provided for the absence of this information. The infrastructure for compliance exists; the information does not. The prevalence of only partial implementation therefore raises questions about how meaningful this transparency is in practice (see Section~\ref{sec:related-work}). Crucially, this classification is restricted to cases where empirical evidence of partial implementation is directly observable. It is distinguished from \textit{No}  where no relevant feature or attempt at implementation exists. For all entries classified as \textit{Partially}, we provide an explanation in Appendix~A.1. %\ref{app:partially}.

\subsubsection{No Access}
\textit{No Access} denotes the classification applied when a criterion cannot be empirically evaluated because the necessary information is not publicly observable or technically verifiable by an external reviewer. Importantly, this classification does not imply non-compliance, but rather reflects that the required information remains inaccessible to external auditors.
In practice, this category arises in two main contexts. First, it applies to internal system architecture such as encryption protocols, authentication mechanisms, or identity and access management procedures. For these, neither platform provides publicly accessible technical documentation. Second, it applies to criteria that presuppose external institutional infrastructure, such as interoperability with an up-to-now unavailable EU-wide AI database, which falls outside the scope of any individual platform's current capabilities. The entries labeled \textit{No Access} are described in detail in Appendix~A.2. %\ref{app:noaccess}.

\subsection{Step 4: Evaluation}\label{sec:evaluation}
In the fourth step, we relate the item-level findings from Step 3 to the main and secondary (sub)goals for German transparency registers outlined by~\citet[Table~39]{algorithmwatch}. 
For each subgoal, we aggregate the classifications of all corresponding checklist items and visualize the distribution across the four categories using a color-coded pie-chart: \colorbox{green!30}{Yes}, \colorbox{red!30}{No}, \colorbox{yellow!40}{Partially}, and \colorbox{gray!30}{No Access}. Concretely, each checklist item that maps onto a given goal contributes one observation to the distribution. The concrete mapping of which checklists and items contribute to which goals can be found in Appendix~B %\ref{app:checklists}
Table~2.
%\ref{tab:mappingchecktoradar}.
The resulting pie chart thus reflects not only whether a goal is met, but to what extent and with what degree of consistency across the individual criteria that operationalize it. A goal for which most items are classified \textit{Yes} can be considered substantially fulfilled, whereas a goal dominated by \textit{No} or \textit{No Access} points to a structural gap in the platform's implementation.
This aggregation allows us to move beyond item-level compliance and assess the degree to which each platform fulfills the broader functions that a transparency register is expected to serve.
The aggregated results for the main goals are presented in Table~\ref{tab:goals}
and discussed in detail in Section~\ref{sec:findings}.  A presentation of the secondary goals can be found in Table~10 %\ref{tab:subgoals}
in the Appendix.

\subsection{Step 5: Action Items}
Building on the evaluation in Step 4, we derive a set of platform-specific action items intended to move items currently classified as \textit{No} or \textit{Partially} toward full compliance. Rather than issuing generic recommendations, each action item is grounded directly in the gap analysis produced by Steps 3 and 4 and is tailored to the specific deficiencies identified for MaKI and PLS, respectively. Where Step 4 establishes \textit{that} a gap exists, Step 5 addresses \textit{how} it might be closed.
The resulting recommendations serve a dual purpose. For platform operators, they constitute an actionable roadmap: each item identifies an addressable deficiency and specifies the change required to resolve it. For policymakers and civil society actors, the action items collectively function as a diagnostic instrument, surfacing the structural patterns, such as a recurring absence of technical metadata or a systematic gap in risk assessment, that individual item-level findings alone might obscure.
It should be noted that the action items are bounded by the information available through the methodology described in Steps 1-4. We do not focus on deficiencies which stem from upstream factors, such as the absence of a mandatory reporting obligation or the lack of standardized taxonomies across the German public sector.
The action items are presented alongside the evaluation results in Table~\ref{tab:goals}
and elaborated upon in the next section. 
A complete list is also available in the Appendix~D. %\ref{subsec:actionitems}.

\newcommand{\actionlist}[1]{%
  \begin{minipage}{2.7cm}
    \begin{itemize}[noitemsep, topsep=0pt, leftmargin=*]
      #1
    \end{itemize}
  \end{minipage}%
}

\newcommand{\actionitem}[1]{%
  \begin{minipage}{2.7cm}#1\end{minipage}%
}

\newcommand{\subgoal}[1]{%
  \begin{minipage}{7.8cm}\setlength{\leftskip}{1em}#1\end{minipage}%
}

\newcommand{\pie}[5]{%
  \piechart{#1}{#2}{#3}{#4}{#5}%
}

% Usage: \piechartraw{radius}{n}{yes}{partially}{no}{no_access}
\newcommand{\piechartraw}[6]{%
  \pgfmathsetmacro{\grnPct}{#3/#2*100}
  \pgfmathsetmacro{\yelPct}{#4/#2*100}
  \pgfmathsetmacro{\redPct}{#5/#2*100}
  \pgfmathsetmacro{\gryPct}{#6/#2*100}
  \piechart{#1}{\grnPct}{\yelPct}{\redPct}{\gryPct}
}

\newcommand{\pieraw}[6]{%
  \begin{minipage}[c]{1cm}\piechartraw{#1}{#2}{#3}{#4}{#5}{#6}\end{minipage}%
}

\newcolumntype{P}[1]{>{\raggedright\hyphenpenalty=10000\exhyphenpenalty=10000}m{#1}}

\begin{table*}[!htp]
  \centering
  \setlength{\tabcolsep}{4pt}
  \renewcommand{\arraystretch}{2.2}
\begin{tabular}{P{8cm} m{2.7cm} >{\centering}m{1.4cm} @{\hspace{4pt}} | @{\hspace{4pt}} >{\centering}m{1.4cm} m{2.7cm}}
    \toprule 
   % \cmidrule(lr){2-3} \cmidrule(lr){4-5}
    \textbf{Goals}
      & \multicolumn{2}{c}{\textbf{MaKI}}
      & \multicolumn{2}{c}{\textbf{PLS}} \\
    \midrule \midrule
  \textbf{1: Ensuring transparency regarding deployed AI systems} & & & & \\
    \subgoal{1a: Publication of detailed information on purpose, function, benefits, risks,
      legal basis, and responsible institution}
      & \actionitem{MA1-3, MA7, B1-3}
      & \raisebox{-0.4\height}{\pie{0.6}{76}{6}{15}{3}}
      & \raisebox{-0.4\height}{\pie{0.6}{70}{0}{30}{0}}
      & \actionitem{LS1-7, B1-3} \\
    \subgoal{1b: Automated and manual review of the submitted data for consistency, completeness, clarity, and timeliness}
      & \actionitem{B13, B14}
      & \raisebox{-0.4\height}{\pie{0.6}{20}{0}{0}{80}}
      & \raisebox{-0.4\height}{\pie{0.6}{30}{20}{0}{50}}
      & \actionitem{B13, B14, LS9} \\
    \subgoal{1c: Provision of easily accessible, understandable data for the public and capacity-building for its use}
      & \actionitem{B4-12}
      & \raisebox{-0.4\height}{\pie{0.6}{48}{19}{24}{9}}
      & \raisebox{-0.4\height}{\pie{0.6}{48}{24}{24}{4}}
      & \actionitem{B4-12, LS9} \\
    \midrule
  \textbf{2: Strengthening trust in the work of public administration} & & & & \\
    \subgoal{2a: Identification of fields of application, criticality, and responsible actors}
      & \actionitem{MA1-3, MA7, B1-3, B16}
      & \raisebox{-0.4\height}{\pie{0.6}{77}{3}{17}{3}}
      & \raisebox{-0.4\height}{\pie{0.6}{66}{3}{31}{0}}
      & \actionitem{LS1-7, LS9, B1-3, B16} \\
    \midrule
  \textbf{3: Enabling dialogue and oversight} & & & & \\
    \subgoal{3a: Publication of reports to support citizen participation and implementation of feedback mechanisms for citizens and organizations}
      & \actionitem{MA5, MA6, B11, B12, B14, B15, B17, B18}
      & \raisebox{-0.4\height}{\pie{0.6}{31}{0}{23}{46}}
      & \raisebox{-0.4\height}{\pie{0.6}{46}{8}{23}{23}}
      & \actionitem{B11, B12, B14, B15, B17, B18} \\
      \subgoal{3b: Implementation of traceability and reporting obligations}
      & \actionitem{MA5, MA6, B11, B12, B14, B16-18}
      & \raisebox{-0.4\height}{\pie{0.6}{12}{0}{18}{70}}
      & \raisebox{-0.4\height}{\pie{0.6}{17.5}{12}{17.5}{53}}
      & \actionitem{LS8, LS9, B11, B12, B14, B16-18} \\
    \subgoal{3c: Awareness-raising through network building, citizen forums, and long-term partnerships}
      & \actionitem{B4-8}
      & \raisebox{-0.4\height}{\pie{0.6}{55}{9}{36}{0}}
      & \raisebox{-0.4\height}{\pie{0.6}{55}{9}{36}{0}}
      & \actionitem{B4-8} \\
      \midrule
  \textbf{4: Centralization of information on AI systems} & & & & \\
    \subgoal{4a: Establishment of a central register documenting all AI systems}
      & \actionitem{B13}
      & \raisebox{-0.4\height}{\pie{0.6}{50}{0}{0}{50}}
      & \raisebox{-0.4\height}{\pie{0.6}{0}{50}{0}{50}}
      & \actionitem{B13, LS9} \\
    \subgoal{4b: Development of an interoperable architecture with automated interfaces}
      & \actionitem{B13}
      & \raisebox{-0.4\height}{\pie{0.6}{50}{0}{0}{50}}
      & \raisebox{-0.4\height}{\pie{0.6}{0}{50}{0}{50}}
      & \actionitem{B13, LS9} \\
      \midrule
  \textbf{5: Ensuring interoperability with the EU database} & & & & \\
    \subgoal{5a: Implementation of an automated interface for communication with the EU database}
      & \actionitem{MA1-3, MA7, B1-3, B13, B16}
      & \raisebox{-0.4\height}{\pie{0.6}{73}{3}{18}{6}}
      & \raisebox{-0.4\height}{\pie{0.6}{61}{3}{33}{3}}
      & \actionitem{LS1-7, LS9, B1-3, B13, B16} \\
    \subgoal{5b: Ensuring regular data updates and synchronization}
      & \actionitem{B13, B16, B19}
      & \raisebox{-0.4\height}{\pie{0.6}{25}{0}{25}{50}}
      & \raisebox{-0.4\height}{\pie{0.6}{0}{25}{25}{50}}
      & \actionitem{LS9, B13, B16, B19} \\
    \subgoal{5c: Substantive consistency of information fields with the EU database and cooperation with EU supervisory authorities}
      & \actionitem{MA1-3, MA7, B1-3, B13}
      & \raisebox{-0.4\height}{\pie{0.6}{73}{3}{12}{12}}
      & \raisebox{-0.4\height}{\pie{0.6}{63}{3}{28}{6}}
      & \actionitem{LS1-7, LS9, B1-3, B13} \\
      \bottomrule
  \end{tabular}
  \caption{Primary goals and subgoals of an AI register, how MaKI and PLS fulfill them, and the action items needed to meet them.}
  \label{tab:goals}
\end{table*}

\section{Findings} \label{sec:findings}
We present our compliance visualization of the primary goals in Table~\ref{tab:goals}. 
The visualization is structured so that pie charts are available
for all primary goals and subgoals of a register as conceptualized in this paper and by \citet{algorithmwatch}.
The rationale for this presentation is to provide a compact
overview of how many checklist items are fulfilled, 
partially fulfilled, not fulfilled or currently not 
assessable due to our access limitations as external auditors.

We note that Table~\ref{tab:goals} does not provide a comparative weighting of the checklist items. 
Thus, the resulting visualization should not be taken as a complete assessment
of the level of fulfillment of certain goals.
To give a concrete example, two of the unfulfilled checklist items in subgoal 1a (Publication of detailed information on purpose, function, benefits, risks, legal basis, and responsible institution) for MaKI are i) to offer the responsible institution's logo
on the AI system's description page; ii) discussing the fundamental rights impact assessment results as per Article 27(1)(a)–(f) of the EU AI Act. 
Clearly, these two points do not warrant the same urgency, but take equal space in the visualization in Table~\ref{tab:goals}. 
The purpose of the present section and analysis will therefore be to further qualify the data we extracted
so to present the main results of our audit.

\paragraph{Goal 1: Ensuring transparency regarding deployed AI systems}

In terms of the published basic information about the system (\textbf{Goal 1a}), MaKI and
PLS perform reasonably well; the picture differs
when one moves onto information about criticality and risks of
the system. PLS does not publish any information
about the risk classification of the system, the risk
assessment or identified risks. MaKI does have this
information, but the treatment of risk assessment is minimal, offering a classification but no further details.
Compared to Lorenz's proposal, a URL to a document providing
concrete, detailed information about risk assessment
methodology employed and the result of the analysis is not included.
In terms of whether the two platforms review the submitted data (\textbf{Goal 1b}), we can provide only a limited assessment for MaKI, as detailed information is not available to external users, nor were we able to assess the data entry forms, as these are not public.
For PLS, an input form is available and we were therefore able to assess some capabilities for automated validation (see Appendix~ A.1). %\ref{app:partially}.
However, the overall procedures for data quality assurance are not visible to us.
In particular, we are not able to assess whether the data is updated every few months and whether the systems appear soon after their commissioning, or whether an update to the data is required after substantial changes. 
Lastly, both platforms have some ways to support the creation of understandable data for the public (\textbf{Goal 1c}). Here, MaKI's exporting capabilities and search functionalities stand out, while PLS offers a map-based visualization of which systems are active in different states as a landing page of sorts.
Capacity-building is perhaps the main weakness here, as there are only limited feedback mechanisms for both platforms.
Compared to our checklist, there is no citizen forum or an equivalent dialogue channel by which the public may be able to ask questions, raise concerns or provide feedback, nor to signal problems with specific data entries.

\paragraph{Goal 2: Strengthening trust in the work of the public administration}
We observe here that, at least in name, both MaKI and PLS allow public institutions at the federal, state, and municipal levels to publish information on AI systems, which supports institutional visibility and accountability. MaKI appears somewhat stronger in this regard because it is specifically focused on AI systems used in public administration and therefore aligns more closely with the objective of administrative transparency. However, the platform is not open to external input, which limits broader participation and independent scrutiny. In addition, both initiatives provide only limited information on important trust-related aspects such as risk assessments, data protection impact assessments, and risk mitigation measures. This weakens the ability of citizens and affected groups to evaluate whether AI systems are being deployed responsibly and in compliance with legal safeguards. Finally, stronger alignment with standardized EU reporting structures and the future EU database is needed to improve comparability, consistency, and trustworthiness across public-sector AI registers.

\paragraph{Goal 3: Enabling dialogue and oversight} Regarding \textbf{Goal 3a}, both MaKI and PLS provide publicly accessible system profiles and geographical visualizations that help citizens identify AI systems active in their municipality or state. However, both initiatives lack sufficiently developed feedback mechanisms that would allow citizens or organizations to submit structured feedback on individual entries or on the platforms as a whole. In addition, the platforms do not explicitly address diverse or critical audiences, and more in-depth information on AI systems, such as training data or validation procedures, is generally not publicly available. Concerning \textbf{Goal 3b}, the implementation of reporting obligations and traceability mechanisms remains partial and difficult to evaluate from the perspective of external auditors. PLS appears more transparent regarding submission structures, whereas MaKI restricts data entry to public administration actors. However, both initiatives provide limited public information on validation processes, quality assurance, database architecture, and update procedures, and the reporting model currently remains voluntary. With respect to \textbf{Goal 3c}, one positive aspect is the establishment of collaborations with educational and civil society organizations and the availability of examples of best practices. However, to the best of our knowledge, no guides or tutorials for using the registers are available, nor have workshops or seminars on reporting been organized. More broadly, there is little evidence of participatory structures such as citizen forums, long-term partnerships, or independent advisory boards integrating ethical, legal, societal, and technical perspectives. Overall, the absence of the AI supervisory authority (\textit{KI-Aufsichtsbehörde}) envisioned by~\citet{algorithmwatch} makes institutional responsibilities difficult to assess, but this only increases the importance of transparent reporting and participation mechanisms within the platforms.\footnote{Researchers replicating this analysis for other EU member states should verify whether an equivalent institution exists and whether it has proposed a concrete governance framework.} 
\paragraph{Goal 4: Centralization of information on AI systems}
Both MaKI and PLS provide web-based interfaces and appear to possess the technical infrastructure necessary to support this objective. At the same time, the assessment of both initiatives remains limited because there is insufficient public information about their database architecture, synchronization interfaces, and quality assurance processes. MaKI is more closely aligned with the requirements of a formal transparency register because it uses standardized fields corresponding to Annex VIII of the AI Act, making it more suitable for structured reporting and legal documentation. In contrast, PLS offers a broader overview of the AI ecosystem, but its data structure is less clearly aligned with the reporting requirements of Annex VIII, particularly regarding conformity assessment documentation such as the EU Declaration of Conformity. 

\paragraph{Goal 5: Ensuring interoperability with the EU database}
Although Article 71 of the AI Act provides for an EU AI database, it was not publicly operational at the time of our evaluation, making the assessment of interoperability largely prospective. Regarding \textbf{Goal 5a}, neither MaKI nor PLS currently provides clearly documented automated synchronization interfaces with the future EU database or with external analysis and reporting tools. However, MaKI appears better positioned in this regard, as mentioned above, because its standardized information fields are more closely aligned with the requirements of the AI Act and Annex VIII. Concerning \textbf{Goal 5b}, it also remains unclear whether either initiative supports versioning or bitemporal historization that would make changes to AI system entries traceable over time, including updates, corrections, or changes in deployment and responsibility. With respect to \textbf{Goal 5c}, cooperation and synchronization with EU supervisory authorities are not yet established, and the assessment remains limited because the relevant technical details and governance structures are not publicly accessible and supervisory responsibilities are themselves still emerging.

\section{Action Needed and Discussion}

Summarizing the preceding section, we find that the goals of a register are only partially fulfilled by the two platforms we evaluated.
We give in Appendix~D %\ref{subsec:actionitems}
a complete list of the actions needed for these platforms to fulfill the objectives of a register; in the following, we touch upon the most urgent and limiting issues we find in MaKI and PLS.

\noindent\textbf{Transparency towards the public is not the core mission.} MaKI's self-described mission as a ``marketplace for AI opportunities" undermines its credibility as a neutral transparency instrument. PLS, while broader in scope, similarly lacks a clear mandate to serve as a register upon which accountability claims can be grounded. For either initiative to function as a national transparency register, the primary and overarching goals must be transparency toward the public. Crucially, such a register should be developed with the people it is meant to serve, through participatory processes such as user surveys or public consultations. We find no evidence that either MaKI or PLS has undertaken such an effort, or is planned for the future. 

\noindent\textbf{Lack of information on risk and impact assessments.} Both registers lack detailed fields for risk and impact assessments. Since neither platform is oriented towards public-facing transparency, serving instead as a promotional instrument for government innovation, any future effort to reach a critical audience~\cite{kemper_transparent_2019} will need to collect this information anew, duplicating work that could have been captured at registration. 
MaKI offers a minimal risk categorization for some systems, but provides no concrete description of oversight measures or actions to be taken should risks materialize. The current registers therefore offer limited transparency (Goal~1) and cannot meaningfully strengthen trust in public administration (Goal~2). 

\noindent\textbf{Quality of data entries.}
Although MaKI uses standardized data formats listed in Annex VIII of the AI Act more consistently than PLS, the content of certain information fields remains questionable. Categories within the MaKI field “AI Method” combine fundamentally different conceptual levels, including methodologies, application domains, and adjacent technical disciplines. Particularly ambiguous classifications such as “Semantic AI,” “Statistical Modeling,” and “Information Retrieval” illustrate the absence of proper terminology. This undermines the objective of an algorithm register because the purpose of such a register is not merely to label systems as “AI,” but to make them interpretable and comparable across institutions and use cases. This may also indicate insufficient quality assurance mechanisms for data entered by reporting authorities and organizations.

\noindent\textbf{Lack of concrete feedback mechanisms.} Neither platform offers citizens a way to give feedback on registered systems or pose questions. There is no public dialogue support (e.g.\ comment fields), no user satisfaction surveys, and no clear consultation opportunities for affected citizens. This impacts Goals 2 and 3, ``Strengthening trust'' and ``Enabling dialogue and oversight.'' 

\noindent\textbf{Public documentation.} Neither platform publishes information on its data collection and quality assurance processes, such as entry deadlines, update obligations, or review steps, nor does it provide user guides for navigating and interpreting register entries. 
MaKI offers export functionality for systematic analysis, but neither platform provides API access.  

\noindent\textbf{Coverage.} We were unable to confirm whether all relevant AI systems, or only a selection, are documented on MaKI and PLS (Goal~4). 
Echoing~\citet{cath_dutch_2022}, we note the absence of any systems published by the Federal Criminal Police (Bundeskriminalamt) or the Federal Police Office (Bundespolizeiamt), despite credible reports of AI usage in law enforcement\footnote{See e.g.\ \url{https://www.heise.de/en/news/NRW-police-modernizes-data-analysis-Who-meets-the-criteria-besides-Palantir-11273481.html}}. Furthermore, in MaKI only AI systems used by the public administration are registered, limiting the coverage by design. We believe that neither initiative operates on a compulsory basis; both rely on voluntary participation, whereas a national register would require well-defined criteria specifying which systems must be registered and which are exempt. 

\section{Conclusion} This paper examined MaKI and PLS as the two main German initiatives for registering AI systems. Our analysis shows that both contribute to the transparency of AI systems, but only partially fulfill the functions expected from a meaningful national transparency register. %Simply presenting information about AI systems or calling a platform a "transparency register" does not, by itself, create transparency.
The central question is not which information a transparency register claims to capture but whether this information is actually present in the register, how understandable it is, and whether the public can meaningfully use it. Germany currently lacks a centralized national AI transparency register. Existing initiatives remain fragmented, differ substantially in purpose and structure, and appear oriented toward internal administrative documentation rather than meaningful transparency for the public. As a result, important elements such as risk and impact assessments, documentation quality, public explanations, feedback mechanisms, identifiable contact persons for inquiries, and accountability for the accuracy of entries remain underdeveloped. Rather than creating additional disconnected initiatives, future efforts should focus on consolidating and strengthening existing ones, complemented by interviews with operators, administrators, and users to capture practical experiences not visible from the register content alone, and by validating the checklists applied here against other national and international AI registration initiatives. Combining the strongest elements of MaKI and PLS could provide the foundation for a national transparency register that is coherent, publicly usable, and aligned with emerging European standards, serving not only public administrations but also citizens, researchers, universities, and other actors seeking reliable information about AI systems beyond the high-risk applications covered by the EU AI Act database. Although no harmonized European standard for AI transparency registers currently exists, Germany has the opportunity to establish a national model for transparent and trustworthy AI governance that is genuinely informative for the public rather than merely administrative in form.

\section*{Acknowledgements}
IP, XH and MC were supported by the ``TOPML: Trading Off Non-Functional Properties of Machine Learning'' project funded by Carl Zeiss Foundation, grant number P2021-02-014.
\bibliography{references}

@misc{maki,
  author       = {{Das vom BMI initiierte Beratungszentrum für K{\"u}nstliche Intelligenz (BeKI)}},
  title        = {Marktplatz der KI-M{\"o}glichkeiten (MaKI): Startseite},
  year         = {2025},
  note         = {\url{https://maki.beki.bund.de/a/bmi-makimo-app?kiosk}, Accessed April 2026}
}

@misc{algorithmwatch,
  author       = {Alina Lorenz},
  title        = {Konzept zur fachlichen, organisatorischen und technischen Umsetzung eines nationalen KI-Transparenzregisters},
  year         = {2025},
  note         = {\url{https://algorithmwatch.org/de/wp-content/uploads/2025/05/2505-AW_KI-TR-Konzept.pdf}, Accessed April 2026}
}

@legal{AIA2024,
  author = {{European Parliament and Council}},
  title = {Regulation (EU) 2024/1689 (Artificial Intelligence Act)},
  journal = {Official Journal of the European Union},
  year = {2024},
  volume = {L 2024/1689},
  url = {http://data.europa.eu/eli/reg/2024/1689/oj},
  note = {Accessed: March 2026}
}

@techreport{david_roth-isigkeit_stellungnahme_2024,
	title = {Stellungnahme im {Rahmen} der öffentlichen mündlichen {Anhörung} am 15. {Mai} 2024 des {Digitalausschusses} des {Bundestags} zum {Thema} "{Nationale} {Spielräume} bei der {Umsetzung} des europäischen {Gesetzes} über künstliche {Intelligenz}"},
	url = {https://www.bundestag.de/resource/blob/1002542/9399017597afda26d626a23617a30bbb/Roth-Isigkeit.pdf},
	urldate = {2026-03-18},
	author = {{David Roth-Isigkeit}},
	month = may,
	year = {2024}
}

@techreport{soizic_penicaud_making_2025,
	title = {Making {Algorithm} {Registers} {Work} for {Meaningful} {Transparency}},
	url = {https://iaciudadana.org/2025/03/13/making-algorithm-registers-work-for-meaningful-transparency/},
	urldate = {2026-03-17},
	author = {{Soizic Pénicaud} and {IA Ciudadana}},
	month = mar,
	year = {2025},
}

@online{noauthor_algorithm_nodate,
    author = {{Eurocities Network}},
	title = {Algorithm {Register} - {Algorithmic} {Transparency} {Standard}},
	url = {https://www.algorithmregister.org/standard},
	urldate = {2026-03-17},
    year = {2026}
}

@article{floridi_artificial_2020,
	title = {Artificial {Intelligence} as a {Public} {Service}: {Learning} from {Amsterdam} and {Helsinki}},
	volume = {33},
	issn = {2210-5441},
	shorttitle = {Artificial {Intelligence} as a {Public} {Service}},
	url = {https://doi.org/10.1007/s13347-020-00434-3},
	doi = {10.1007/s13347-020-00434-3},
	language = {en},
	number = {4},
	urldate = {2026-03-17},
	journal = {Philosophy \& Technology},
	author = {Floridi, Luciano},
	month = dec,
	year = {2020},
	pages = {541--546},
}

@article{cath_dutch_2022,
	title = {Dutch {Comfort}: {The} {Limits} of {Ai} {Governance} {Through} {Municipal} {Registers}},
	volume = {26},
	shorttitle = {Dutch {Comfort}},
	doi = {10.5840/techne202323172},
	number = {3},
	journal = {Techné Research in Philosophy and Technology},
	author = {Cath, Corinne and Jansen, Fieke},
	year = {2022},
	pages = {395--412},
}

@article{nieuwenhuizen_algorithm_2024,
	title = {Algorithm {Registers}: {A} {Box}-{Ticking} {Exercise} or {Meaningful} {Tool} for {Transparency}?},
	volume = {29},
	url = {https://journals.sagepub.com/doi/10.1177/15701255241297107},
	doi = {10.1177/15701255241297107},
	number = {4},
	urldate = {2026-03-17},
	journal = {Information Polity},
	author = {Nieuwenhuizen, Esther},
	year = {2024},
}

@article{kemper_transparent_2019,
	title = {Transparent to whom? {No} algorithmic accountability without a critical audience},
	volume = {22},
	issn = {1369-118X},
	shorttitle = {Transparent to whom?},
	url = {https://doi.org/10.1080/1369118X.2018.1477967},
	doi = {10.1080/1369118X.2018.1477967},
	number = {14},
	urldate = {2026-03-17},
	journal = {Information, Communication \& Society},
	publisher = {Routledge},
	author = {Kemper, Jakko and Kolkman, Daan},
	month = dec,
	year = {2019},
	note = {\_eprint: https://doi.org/10.1080/1369118X.2018.1477967},
	pages = {2081--2096},
}

@misc{open_government_partnership_promoting_2025,
	title = {Promoting {Government} {Transparency} of {Algorithmic} {Tools} in the {United} {Kingdom}},
	url = {https://www.opengovpartnership.org/united-kingdom-digital-governance-story/},
	author = {{Open Government Partnership}},
	year = {2025},
}

@techreport{leslie_securing_2024,
	title = {Securing meaningful transparency of public sector use of {AI}: {Comparative} approaches across five jurisdictions},
	url = {https://publiclawproject.org.uk/content/uploads/2024/10/Securing-meaningful-transparency-of-public-sector-AI.pdf},
	institution = {Public Law Project},
	author = {Leslie, M. and Selman, C.},
	year = {2024},
}

@techreport{global_partnership_on_artificial_intelligence_algorithmic_2024,
	title = {Algorithmic {Transparency} in the {Public} {Sector}: {A} state-of-the-art report of algorithmic transparency instruments},
	url = {https://wp.oecd.ai/app/uploads/2025/05/algorithmic-transparency-in-the-public-sector.pdf},
	author = {{Global Partnership on Artificial Intelligence}},
	year = {2024},
}

@techreport{ada_lovelace_institute_algorithmic_2021,
	title = {Algorithmic {Accountability} for the {Public} {Sector}: {Learning} from the {First} {Wave} of {Policy} {Implementation}},
	url = {https://www.adalovelaceinstitute.org/report/algorithmic-accountability-public-sector/},
	author = {{Ada Lovelace Institute} and {AI Now Institute} and {Open Government Partnership}},
	year = {2021},
}

@techreport{haataja_public_2020,
	title = {Public {AI} {Registers}: {Realising} {AI} transparency and civic participation in government use of {AI}},
	url = {https://ai.hel.fi/wp-content/uploads/White-Paper.pdf},
	institution = {Saidot},
	author = {Haataja, Meeri and van de Fliert, Linda and Rautio, Pasi},
	month = sep,
	year = {2020},
}

@inproceedings{mitchell_model_2019,
	title = {Model {Cards} for {Model} {Reporting}},
	doi = {10.1145/3287560.3287596},
	booktitle = {Proceedings of the {Conference} on {Fairness}, {Accountability}, and {Transparency}},
	author = {Mitchell, Margaret and Wu, Simone and Zaldivar, Andrew and Barnes, Parker and Vasserman, Lucy and Hutchinson, Ben and Spitzer, Elena and Raji, Inioluwa Deborah and Gebru, Timnit},
	year = {2019},
}

@article{busuioc_reclaiming_2023,
	title = {Reclaiming transparency: contesting the logics of secrecy within the {AI} {Act}},
	volume = {2},
	issn = {2752-6135},
	shorttitle = {Reclaiming transparency},
	url = {https://www.cambridge.org/core/journals/european-law-open/article/reclaiming-transparency-contesting-the-logics-of-secrecy-within-the-ai-act/01B90DB4D042204EED7C4EEF6EEBE7EA},
	doi = {10.1017/elo.2022.47},
	language = {en},
	number = {1},
	urldate = {2026-03-18},
	journal = {European Law Open},
	author = {Busuioc, Madalina and Curtin, Deirdre and Almada, Marco},
	month = mar,
	year = {2023},
	pages = {79--105},
}

@techreport{european_commission_guidelines_2025,
	title = {Guidelines on the definition of an artificial intelligence system established by {Regulation} ({EU}) 2024/1689 ({AI} {Act})},
	url = {https://digital-strategy.ec.europa.eu/en/library/commission-publishes-guidelines-ai-system-definition-facilitate-first-ai-acts-rules-application},
	author = {{European Commission}},
	month = feb,
	year = {2025},
}

@article{bell_think_2023,
	title = {Think about the stakeholders first! {Toward} an algorithmic transparency playbook for regulatory compliance},
	volume = {5},
	doi = {10.1017/dap.2023.8},
	journal = {Data \&\#38; Policy},
	author = {Bell, Andrew and Nov, Oded and Stoyanovich, Julia},
	year = {2023},
	pages = {e12},
}

@article{van_vliet_defining_2024,
	title = {Defining and {Implementing} {Algorithm} {Registers}: {An} {Organizational} {Perspective}},
	shorttitle = {Defining and {Implementing} {Algorithm} {Registers}},
	url = {https://aisel.aisnet.org/ecis2024/track04_impactai/track04_impactai/5},
	journal = {ECIS 2024 Proceedings},
	author = {van Vliet, Martijn and Schuitemaker, Nena and Espana, Sergio and Weerd, Inge van de and Brinkkemper, Sjaak},
	month = jun,
	year = {2024},
}

@article{gebru_datasheets_2021,
	title = {Datasheets for datasets},
	volume = {64},
	issn = {0001-0782},
	url = {https://dl.acm.org/doi/10.1145/3458723},
	doi = {10.1145/3458723},
	number = {12},
	urldate = {2026-03-18},
	journal = {Commun. ACM},
	author = {Gebru, Timnit and Morgenstern, Jamie and Vecchione, Briana and Vaughan, Jennifer Wortman and Wallach, Hanna and III, Hal Daumé and Crawford, Kate},
	month = nov,
	year = {2021},
	pages = {86--92},
}

@phdthesis{murad_2021,
	type = {Thesis},
	title = {Beyond the "{Black} {Box}": {Enabling} {Meaningful} {Transparency} of {Algorithmic} {Decision}-{Making} {Systems} through {Public} {Registers}},
	copyright = {In Copyright - Educational Use Permitted},
	shorttitle = {Beyond the "{Black} {Box}"},
	url = {https://dspace.mit.edu/handle/1721.1/139092},
	language = {en},
	urldate = {2026-03-18},
	school = {Massachusetts Institute of Technology},
	author = {Murad, Maya},
	month = jun,
	year = {2021},
	note = {Accepted: 2022-01-14T14:49:25Z},
}

@misc{minbzk_publicatiestandaard_2024,
  author       = {{Ministerie van Binnenlandse Zaken en Koninkrijksrelaties}},
  title        = {{Algoritmeregister Publicatiestandaard}},
  year         = {2024},
  version      = {1.0.10},
  type         = {MinBZK Standaard, consultatieversie},
  date         = {2024-11-29},
  url          = {https://algoritmes.overheid.nl/en},
  urldate      = {2026-05-21}
}
\clearpage 

\begin{appendix}
\section{Methodology Details}
In this section, we give details on why and when we labeled a requirement with partially and no access.  
    \subsection{Partially Fulfilled}\label{app:partially}
\paragraph{Evaluation of Usability and Accessibility (Table~\ref{tab:architecture})}

The concept paper by~\citet{algorithmwatch} defines usability and accessibility of the input portal broadly, including intuitive navigation, input assistance, screen reader compatibility, mobile optimization, and keyboard operability. As these properties cannot be determined solely from documentation, an exploratory functional interface evaluation was conducted. The evaluation combined manual interaction tests and browser-based technical inspection using developer tools. The goal was not to certify compliance with accessibility standards (e.g., WCAG), but to determine whether the portals plausibly support barrier-free interaction in practice.
The following aspects were systematically examined:

\paragraph{1. Keyboard navigation}
\begin{itemize}
    \item Navigation via Tab, Enter, Space and Arrow keys
    \item Ability to operate filters and controls without mouse input
\end{itemize}

\paragraph{2. Screen reader compatibility (structural accessibility)}
\begin{itemize}
    \item Inspection of semantic HTML structure via browser developer tools
    \item Presence of ARIA attributes and textual labels
    \item Identification of labeled form fields and required-field indicators
\end{itemize}

\paragraph{3. Input assistance}
\begin{itemize}
    \item Required field markings
    \item Immediate validation feedback
    \item Presence of structured input constraints (input masks)
\end{itemize}

For MaKI, input assistance could not be fully tested because write access to the portal is restricted to authorized institutional users. In such cases, only observable interface elements were evaluated.  
For PLS, required fields are marked, but no immediate validation feedback is provided, and arbitrary input can be entered without restriction. No input masks were observed.

\paragraph{4. Mobile optimization}
\begin{itemize}
    \item Responsive layout testing using browser responsive mode
    \item Functional equivalence between desktop and small viewport layouts
\end{itemize}

\paragraph{5. Navigation and functionality}
\begin{itemize}
    \item Search, filtering and sorting functions
    \item General interaction clarity and discoverability of functions
\end{itemize}

The MaKI interface provides an overall intuitive interaction structure. Navigation elements, filtering options, and content presentation are easy to locate and understandable without prior instruction. Keyboard navigation is largely functional, and structural accessibility indicators such as ARIA labels and semantic elements are present in the markup, suggesting compatibility with assistive technologies.
Responsive behavior is implemented: the interface adapts to smaller viewports and remains usable on mobile screen sizes, although it is not designed mobile-first.
Input assistance could only be assessed indirectly because data entry is restricted to authorized institutional users. Required fields are marked, and form semantics are present, indicating guided input, but actual validation behavior and error prevention mechanisms could not be empirically tested.
Overall, the system demonstrates substantial support for accessibility and usability, yet cannot be considered fully compliant due to limited verifiability of input workflows and the broad scope of the accessibility criterion. The requirement is therefore rated as partially fulfilled.

The PLS interface also provides clear navigation, search, filter and sorting functionality and is operable via keyboard navigation. Required fields are marked, which supports guided interaction.
However, structured input constraints are largely absent. Fields accept arbitrary input without format restrictions (no observable input masks), reducing data entry guidance and error prevention.
Consequently, the interface supports basic usability and accessibility. The requirement is therefore classified as partially fulfilled.
\subsubsection{Impact assessments (Table~\ref{tab:informationfields})}

Certain fields required consultation of legal provisions beyond the platform entries themselves. This applies in particular to the data protection impact assessment and the fundamental rights impact assessment.
The transparency-register schema requires information corresponding to a data protection impact assessment under Article 35 GDPR (or Article 27 LED), including an indication of whether such an assessment was conducted and a summary of its results. The MaKI platform provides dedicated fields matching these requirements. However, these fields are not completed for any analyzed AI system, and no justification for the absence of an assessment is publicly available. PLS does not contain comparable fields. Consequently, neither platform provides the required assessment information. Nevertheless, MaKI demonstrates structural alignment with the transparency-register specification, whereas PLS does not.
A similar distinction arises for the fundamental rights impact assessment under Article 27 AI Act. MaKI includes fields anticipating this obligation (e.g., risk description, affected persons, oversight, and mitigation measures), but they are not consistently filled, and no explanation is provided when information is missing. 
Accordingly, these fields are evaluated as partially fulfilled for MaKI; the disclosure structure exists but lacks substantive entries, and not fulfilled for PLS. 

\subsubsection{Standardized data formats (Table~\ref{tab:architecture})}

The concept paper by~\citet{algorithmwatch} requires the use of standardized data formats to ensure comparability and interoperability of entries. To operationalize this requirement, the evaluation was aligned with Annex VIII of the AI Act, which specifies the minimum information to be submitted upon registration of high-risk AI systems under Article 49. Annex VIII defines a structured set of mandatory information fields, including provider identification, system identification, intended purpose, operating logic, system status, and certification-related information such as the EU Declaration of Conformity and notified body references.
The assessment therefore examined whether the platforms provide structured input fields corresponding to the categories defined in Annex VIII.

The MaKI register provides input structures corresponding to the categories specified in Annex VIII. The interface contains predefined fields covering provider identification, system identification, intended purpose, and operational characteristics. Although entries are not always fully completed, the presence of these structured fields demonstrates that the data model itself follows a standardized schema compatible with regulatory reporting requirements.
Yes (standardized format exists), but only partially fulfilled due to lack of explanation for missing entries.

PLS provides structured descriptive fields for AI systems, including purpose and functional description. However, the schema does not systematically reflect the regulatory structure defined in Annex VIII. In particular, no structured references to conformity assessment documentation, such as the EU Declaration of Conformity or notified body certification details, were observed. The requirement is therefore classified as partially fulfilled.
\subsubsection{Integrated feedback functions (Table~\ref{tab:process})}

The proposal by~\citet{algorithmwatch} requires an integrated feedback mechanism that allows users to provide feedback both on individual AI system entries and on the platform itself. According to the conceptual specification of the transparency register, such feedback should be reviewed by the supervisory authority and, where appropriate, forwarded to the responsible institutions.

Both PLS and MaKI provide a feedback option directed at the platform as a whole. However, no clearly identifiable mechanism allows feedback to be attached to individual AI system entries. Furthermore, the internal processing of submitted feedback cannot be evaluated without interacting with administrative procedures.

Consequently, the requirement is considered only partially fulfilled.

\subsection{No Access}\label{app:noaccess}
\subsubsection{Security, Data Protection and Database Interoperability (Table~\ref{tab:architecture})}

The framework by~\citet{algorithmwatch} requires that security and data protection be ensured through compliance with national and European data protection standards, including encrypted transmission, controlled access, and identity access management. Assessing these aspects requires insight into backend infrastructure, authentication mechanisms, and system architecture documentation that is not publicly accessible in either evaluated platform.

Neither platform provides verifiable technical details about encryption standards, authentication procedures, role-based access control implementation, or identity management systems that are publicly documented. As a result, the presence, quality, and correctness of implemented security mechanisms cannot be empirically evaluated. Consequently, this criterion is classified as \emph{no access} for both platforms.
\paragraph{Central and interoperable databases (Table~\ref{tab:process})}

The framework further requires a central database architecture ensuring interoperability with a European Union database and cooperation with competent European bodies. During the evaluation (February 2026), no operational EU database for AI systems was publicly available. Although some entries contain references to a future \textit{EU database entry}, these references could not be resolved to an existing infrastructure.

Furthermore, no publicly identifiable central authority responsible for synchronizing data exchange between the evaluated registers and a European database could be observed. Because interoperability inherently depends on the existence of such an external infrastructure, its implementation cannot be tested in practice.

The absence of a verifiable external database and the lack of observable exchange interfaces lead to a classification of \emph{no access} rather than non-compliance. The evaluation distinguishes between implemented functionality and announced or anticipated functionality; only technically observable features are considered fulfilled criteria.
%mapping from checklists to radar charts

\begin{table*}[!htp]
  \centering
  \setlength{\tabcolsep}{4pt}
  \renewcommand{\arraystretch}{2.2}
\begin{tabular}{p{6cm}|p{6cm}|p{2.3cm}|p{2.8cm}}
    \toprule 
   % \cmidrule(lr){2-3} \cmidrule(lr){4-5}
    Goals
      & Related Checklist Items& Count MaKI (yes, no, partially, no access) & Count PLS (yes, no, partially, no access)\\
    \midrule \midrule
 
    1a: Publication of detailed information on purpose, function, benefits, risks,
      legal basis, and responsible institution
      & Table \ref{tab:informationfields}, Table \ref{tab:scope} (affected systems)& 25,5,2,1&23,10,0,0\\
   1b: Automated and manual review of the submitted data for consistency, completeness, clarity, and timeliness
      & Table~\ref{tab:process} (data collection, QA), Table~\ref{tab:architecture} (DB architecture) & 2,0,0,8&3,0,2,5 \\
    1c: Provision of easily accessible, understandable data for the public and capacity-building for its use
      & Table~\ref{tab:architecture} (input portal, analysis and reporting tool, result presentation), Table~\ref{tab:process} (reporting), Table~\ref{tab:campaigns} & 10,5,4,2&10,5,5,1\\
    \midrule
  
    2a: Identification of fields of application, criticality, and responsible actors
      &Table~\ref{tab:informationfields}, Table \ref{tab:scope} (publishing institutions), Table \ref{tab:process} (Data alignment with EU database), Table~\ref{tab:architecture} (standardized data formats) & 27,6,1,1&23,11,1,0 \\
    \midrule
  
    3a: Publication of reports to support citizen participation and implementation of feedback mechanisms for citizens and organizations
      & Table \ref{tab:process} (feedback process, reporting, QA), Table \ref{tab:reporting} & 4,3,0,6&6,3,1,3 \\
      3b: Implementation of traceability and reporting obligations
      &  Table~\ref{tab:process},Table~\ref{tab:responsibility},Table \ref{tab:architecture} (DB arch.) & 2,3,0,12&3,3,2,9 \\
      3c: Awareness-raising through network building, citizen forums, and long-term partnerships
      & Table~\ref{tab:campaigns}&6,4,1,0&6,4,1,0\\
      \midrule

    4a: Establishment of a central register documenting all AI systems
      & Table \ref{tab:architecture}&1,0,0,1&0,0,1,1 \\
    4b: Development of an interoperable architecture with automated interfaces
      & Table \ref{tab:architecture}&1,0,0,1&0,0,1,1  \\
      \midrule

    5a: Implementation of an automated interface for communication with the EU database
      & Table~\ref{tab:informationfields}, Table~\ref{tab:architecture} (DB architecture), Table \ref{tab:process} (Data alignment with EU database &24,6,1,2&20,11,1,1 \\
    5b: Ensuring regular data updates and synchronization
      & Table~\ref{tab:architecture} (DB architecture, security and data protection measures), Table \ref{tab:process} (Data alignment with EU database &1,1,0,2&0,1,1,2  \\
    5c: Substantive consistency of information fields with the EU database and cooperation with EU supervisory authorities
      &Table~\ref{tab:reporting},Table~\ref{tab:architecture} (DB architecture), Table \ref{tab:responsibility} &24,4,1,4&20,9,1,3 \\
      \bottomrule
  \end{tabular}
  \caption{Main objectives of an AI register and their mapping from the checklists as well as the numbers from which the charts are extracted.}
  \label{tab:mappingchecktoradar}
\end{table*}
\section{Checklists}\label{app:checklists}
We here provide the checklists in table format derived from the proposal by~\citet{algorithmwatch} during Step 2 (Section~\ref{sec:checklist}) of our analysis. The assessment of the two existing platforms MaKI and PLS was carried out against these checklists (Step 3, Section~\ref{sec:assesment}). The mapping from checklists to the radar charts can be found in Table~\ref{tab:mappingchecktoradar}.

Table~\ref{tab:informationfields} presents the set of information fields that a transparency register should capture, ranging from institutional identifiers and system descriptors to risk‑related assessments and contact data, and indicates the extent to which each platform provides the required entries.  The end‑to‑end data‑management life‑cycle is illustrated in Table~\ref{tab:process}, which details the collection, quality‑assurance, alignment with the EU database, reporting and feedback processes.  Table~\ref{tab:architecture} establishes the technical architecture underlying the register, including the web‑based input portal, database design, analytics and reporting tools, and the public result‑presentation layer. The spectrum of user‑oriented outputs is summarized in Table~\ref{tab:reporting}, where basic list presentations, AI profiles, geographic maps and aggregated statistics are listed as well as API access.  To foster awareness and capacity building, Table~\ref{tab:campaigns} enumerates the communication methods, training activities and participatory formats that should be offered to institutions and to the broader public.  Governance and oversight requirements are captured in Table~\ref{tab:responsibility}, which delineates professional, organizational and technical responsibilities as well as the role of an interdisciplinary expert panel, none of which are currently documented for the evaluated registers.  Finally, Table~\ref{tab:scope} defines the intended scope and coverage of the register: mandatory reporting of AI systems as required by the KI‑VO, restriction to operational and non‑classified systems, and inclusion of federal, state and municipal public bodies.

\section{Secondary Goals}\label{app:sbgoals}
Table~\ref{tab:subgoals} visualizes the compliance of MaKI and PLS with the secondary goals in Lorenz's proposal. 

\begin{table*}[!ht]
\label{tab:information-fields}
    \centering
    \caption{List and content of information fields which a transparency register should implement for its entries \cite[chapter 5]{algorithmwatch} as well as the results of our evaluation for MaKI and PLS (Step 3, Section ~\ref{sec:assesment}). } \label{tab:informationfields}
    \begin{tabular}{l|l|l|l}
\toprule
        \textbf{Information Field} & \textbf{Content} & \textbf{MaKI} & \textbf{PLS} \\ \midrule \midrule
        \textbf{Institution} & name & yes & yes \\ 
        \textbf{} & type & yes & yes \\ 
        \textbf{} & further details & yes & yes \\ 
        \textbf{} & address & yes & yes \\ 
        \textbf{} & logo & no & yes \\ 
        \textbf{} & website & yes & yes \\ 
        \textbf{} & role according to KI-VO & yes & yes \\ \midrule
        \textbf{KI system} & name & yes & yes \\
        \textbf{} & link to EU-database entry & no & no \\ 
        \textbf{} & description & yes & yes \\
        \textbf{} & purpose and impact & yes & yes \\ 
        \textbf{} & website & yes & yes \\ 
        \textbf{} & technology areas & yes & yes \\ 
        \textbf{} & application areas & no & yes \\ 
        \textbf{} & added value & yes & yes \\ 
        \textbf{} & used information and operating logic & yes & yes \\ 
        \textbf{} & commissioning date & yes & yes \\ 
        \textbf{} & decommissioning date & yes & yes \\ \midrule
        \textbf{Criticality} & risk classification (KI-VO) & yes & no \\ 
        \textbf{} & risk assessment & no & no \\
        \textbf{} & identified risks & yes & no \\ 
        \textbf{} & risk mitigation measures & yes & no \\ 
        \textbf{} & additional risk assessment information & no & no \\ 
        \textbf{} & data protection impact assessment & yes & no \\ 
        \textbf{} & result of data protection impact assessment & partially & no \\ 
        \textbf{} & fundamental rights impact assessment & yes & no \\ 
        \textbf{} & result of fundamental rights impact assessment & yes & no \\ \midrule
        \textbf{Contact} & contact person & yes & yes \\
        \textbf{} & consent to storage & no access & yes \\ 
        \textbf{} & data protection notice (only internal) & yes & yes \\ \bottomrule
    \end{tabular}
    
\end{table*}

\begin{table*}[!ht]
\label{tab:functional-process}

    \centering
    \caption{ Functional process description and data processing workflow for the transparency register~\cite[chapter 6]{algorithmwatch} as well as the results of our evaluation for MaKI and PLS (Step 3, Section ~\ref{sec:assesment}).} \label{tab:process}
    \begin{tabular}{p{4cm}|p{9cm}|l|l}
    \toprule
   
        \textbf{Process} & \textbf{Description} & \textbf{MaKI} & \textbf{PLS} \\ \midrule \midrule
        \textbf{Data collection } & Conducted in four steps: institution, KI system, criticality, contact person. & yes & yes \\ 
        \textbf{} & First entry no later than 2 weeks after commissioning. & no access & no access \\ 
        \textbf{} & Automatic reminder for update every 4 months. & no access & no access \\ 
        \textbf{} & Manual update for substantial changes. & no access & no access \\ 
        \textbf{} & Final update no later than 4 weeks after decommissioning. & no access & no access \\ \midrule
        \textbf{Quality assurance } & Check for consistency, completeness, comprehensibility and timeliness & no access & partially \\ 
        \textbf{} & Use of selection fields and checkboxes for standardization (automated validation). & no access & yes \\ 
        \textbf{} & Performed immediately after the initial entry by human staff. & no access & yes \\ \midrule
        \textbf{Data alignment with EU database } & Automated interface for synchronization (export / import). & no & no \\ \midrule
        \textbf{Reporting } & Creation of reports for the various target groups. & no & no \\ \midrule
        \textbf{Feedback process } & Establish feedback mechanisms for citizens and organizations & no access & no access \\ 
        \textbf{} & Review of submissions. & no access & no access \\ 
        \textbf{} & Follow up by responsible institutions. & no access & no access \\ 
        \textbf{} & Survey of user satisfaction. & no & no \\ \bottomrule
    \end{tabular}
\end{table*}
\begin{table*}[!ht]
    \centering
    \caption{Technical architecture and its implementation for the transparency register~\cite[chapter 7]{algorithmwatch} as well as the results of our evaluation for MaKI and PLS (Step 3, Section ~\ref{sec:assesment}).} \label{tab:architecture}
    \begin{tabular}{l|p{9cm}|l|l}
\toprule
        \textbf{Technical Architecture} & \textbf{Implementation} & \textbf{MaKI} & \textbf{PLS} \\ \midrule\midrule
        \textbf{Input portal} & Web based & yes & yes \\ 
        \textbf{} & Validation mechanisms to check completeness and consistency via mandatory fields. & no access & partially \\ 
        \textbf{} & User friendly and barrier free. & yes & yes \\ 
        \textbf{} & Security and data protection measures. & no access & no access \\ \midrule
        \textbf{Database architecture} & Central, interoperable databases (test and production). & no access & no access \\ 
        \textbf{} & Standardized data formats. & yes & partially \\\midrule
        \textbf{Analysis and reporting tool} & Data preparation.  & partially & partially \\
        \textbf{} & Design and creation of standardized reports.  & partially & partially \\ 
        \textbf{} & Visualization tools. & yes & yes \\ \midrule
        \textbf{Result presentation} & Publicly accessible platform with search and filter functions. & yes & yes \\ 
        \textbf{} & Integrated feedback functions. & partially & partially \\ \bottomrule
    \end{tabular}
\end{table*}

\begin{table*}[!ht]
\label{tab:reporting-formats}
    \centering
    \caption{Reporting formats and accessible analytics for the transparency register~\cite[chapter 8]{algorithmwatch} as well as the results of our evaluation for MaKI and PLS (Step 3, Section ~\ref{sec:assesment}).} \label{tab:reporting}
    \begin{tabular}{l|p{10cm}|l|l}
    \toprule
        \textbf{Method} & \textbf{Description} & \textbf{MaKI} & \textbf{PLS} \\ \midrule \midrule
        \textbf{List presentations} & Tabular overview of all information fields. Functionalities: filtering, free-text search, sorting, exporting. & yes & yes \\ \midrule
        \textbf{AI profiles (Steckbriefe)} & Summarized views with basic information, application details, temporal aspects and additional data. Functionalities: export and links to further websites. & yes & yes \\ \midrule
        \textbf{AI maps (Landkarten)} & Geographic visualization of all AI systems. Functionalities: interactive filtering by state/region/city; heatmaps for concentration; clickable markers linking to profiles. & yes & yes \\ \midrule
        \textbf{Aggregated statistics} & Quantitative overviews and trend analyses. Functionalities: dashboards visualizing counts, trends and distributions; benchmarking between institutions, regions or sectors. & yes & yes \\\midrule
        \textbf{API} & Access to all information fields; integration into programs and other databases. & no & no \\ \bottomrule
    \end{tabular}
\end{table*}

\begin{table*}[!ht]
\label{tab:information-campaigns}
    \centering
    \caption{Information campaigns, competence building and sensitization measurements sorted by target groups which should be provided by the transparency register~\cite[chapter 9]{algorithmwatch} as well as the results of our evaluation for MaKI and PLS (Step 3, Section ~\ref{sec:assesment}). } \label{tab:campaigns}
    \begin{tabular}{l|l|l|l}
\toprule
        \textbf{Target  Group} & \textbf{Method} & \textbf{MaKI} & \textbf{PLS} \\ \midrule \midrule
        \textbf{For institutions} & Information campaigns (guides, infographics, videos). & yes & yes \\ 
        \textbf{} & Workshops and webinars on correct reporting. & no & no \\ 
        \textbf{} & Hotlines and FAQs. & partially & partially \\ 
        \textbf{} & Best practice examples. & yes & yes \\ 
        \textbf{} & Promotion of exchange between institutions. & yes & yes \\
        \textbf{} & Guidelines and templates & no & no \\ \midrule
        \textbf{For users} & Understandable reporting (AI maps, profiles). & yes & yes \\ 
        \textbf{} & Information campaigns (media campaigns, infographics). & yes & yes \\ 
        \textbf{} & Guides and tutorials for navigation and analysis. & no & no \\ 
        \textbf{} & Citizens'  forums for dialogue.  & no & no \\ 
        \textbf{} & Long-term partnerships with educational and civil society organizations. & yes & yes \\ \bottomrule
    \end{tabular}
\end{table*}

\begin{table*}[!ht]
\label{tab:professional-responsibility}
    \centering
    \caption{ Professional responsibility, steering and supervision ~\cite[chapter 10]{algorithmwatch} as well as the results of our evaluation for MaKI and PLS (Step 3, Section ~\ref{sec:assesment}).}\label{tab:responsibility}
    \begin{tabular}{l|p{9cm}|l|l}
    \toprule
        \textbf{Responsibility}& \textbf{Requirements} & \textbf{MaKI} & \textbf{PLS} \\ \midrule\midrule
        \textbf{Professional responsibility} &  Implementation and monitoring of reporting obligations; design, implementation, support, monitoring, and continuous improvement of all operational functional processes& no access & no access \\ \midrule
        \textbf{Organizational responsibility} & Coordination of tasks and stakeholders; collaboration with national and European actors; information, capacity building, and awareness raising; resource management & no access & no access \\ \midrule
        \textbf{Technical responsibility} & Operation of the technical infrastructure; IT security and data protection; ensuring interoperability with the EU database. & no access & no access \\ \midrule
        \textbf{Interdisciplinary expert panel} & Advisory board comprising administration, science, civil society and industry; integration of ethical, legal, societal and technical perspectives; independence to guarantee transparency and trust. & no access & no access \\ \bottomrule
    \end{tabular}
\end{table*}

\begin{table*}[!ht]
\label{tab:scope-coverage}
    \centering
    \caption{Scope and coverage of the transparency register~\cite[chapter 4]{algorithmwatch} as well as the results of our evaluation for MaKI and PLS (Step 3, Section ~\ref{sec:assesment}).}\label{tab:scope}
    \begin{tabular}{l|l|l|l}
 \toprule
        \textbf{Scope} & \textbf{Coverage}& \textbf{MaKI} & \textbf{PLS} \\ \midrule \midrule
        \textbf{Affected systems}&AI systems with mandatory reporting according to the KI VO&yes & yes \\ 
         &Restriction to AI systems in operation & partially & yes \\ 
        &Restriction to non-classified AI systems & yes & yes \\ \midrule
         \textbf{Publishing institutions}&Federal-level public institutions& yes & yes \\ 
        &State-level public institutions & yes & yes \\ 
        &Municipal-level public institutions& yes & yes \\ \bottomrule
    \end{tabular}
\end{table*}

\section{Action Items}\label{subsec:actionitems}

\paragraph{Action items derived directly from Table~X.}
The action items are derived on a one-to-one basis from all table entries that are not fully fulfilled; however, their substantive importance differs and is therefore discussed separately in the text.
\paragraph{MaKI.}
\begin{itemize}
    \item[MA1] Clarify the status of the consent-to-storage field or indicate  that this information is not publicly accessible.
   \item[MA2] Add the missing application area information.
   \item[MA3] Add the institution logo where missing.
    \item[MA4] Expand the scope beyond AI systems used in public administration by including AI systems used by companies and research institutions, or by creating a clearly linked reporting pathway for these actors.
    \item[MA5] Disclose whether MaKI uses selection fields, checkboxes, or other structured input mechanisms for standardization and automated validation during data entry.
    \item[MA6] Disclose whether MaKI includes a human quality assurance review after the initial entry, including who performs the review and which criteria are checked.
    \item[MA7] Add the result of data protection impact assessment or a reason why it was not conducted or included.
\end{itemize}

\paragraph{PLS.}
\begin{itemize}
    \item[LS1] Add structured information on risk classification.
    \item[LS2] Add structured information on identified risks.
    \item[LS3] Add structured information on risk mitigation measures.
    \item[LS4] Add a field for data protection impact assessment information.
    \item[LS5] Add a field for the result of the data protection impact assessment.
    \item[LS6] Add a field for fundamental-rights impact assessment information.
    \item[LS7] Add a field for the result of the fundamental-rights impact assessment.
    \item[LS8] Add visible structured input masks and validation mechanisms, including mandatory fields, completeness checks, and consistency checks.
    \item[LS9] Add the missing information fields required under Annex VIII, Sections A--C of the AI Act, hence aligning it with the expected structure of the EU AI database.

\end{itemize}
\paragraph{Both}
\begin{itemize}
    \item[B1] Add a dedicated risk assessment field.
    \item[B2] Add additional risk assessment information.
    \item[B3] Add a field for the link to the EU database entry once such entries become available.
    \item[B4] Add user guides and tutorials for navigation and analysis, including practical instructions on how to search the register, understand individual entries, interpret maps/profiles, and use visualizations. 
    \item[B5] Add interactive learning formats for the public, such as short explanatory videos, walkthroughs, or webinars on how to use the register and interpret its information.
    \item[B6] Add a structured public dialogue platform for questions, concerns, and feedback from citizens, with moderation and clear responsibilities for follow-up.
    \item[B7] Expand understandable reporting by adding more accessible summaries, visual explanations, and plain-language descriptions of register entries for non-expert users.
    \item[B8] Add regular consultation or Q\&A formats with citizens, experts, and public-sector actors to support exchange and build public capacity to use the register.
     \item[B9] Strengthen data preparation functions by making the process for preparing register data for analysis more explicit and reusable.
    \item[B10] Strengthen standardized reporting functions by providing systematic report formats, reusable summaries, and clearer output structures.
    \item[B11] Add feedback functions, allowing users to provide comments, corrections, or concerns directly on individual AI system entries.
    \item[B12] Create standardized reports for different target groups, such as citizens and affected persons, public authorities and institutions, government agencies, civil society and academia, and companies and organizations.
    \item[B13] Publish or document the database architecture, including data models, field definitions, validation rules, and export formats; where possible, make relevant components open source to improve transparency and verifiability.
    \item[B14] Publish concise procedural documentation on the register's data collection and quality assurance process, including entry deadlines, update obligations, reminder mechanisms, review steps, and responsibilities, without requiring public access to the internal input system.
    \item[B15] Provide a public API with access to all register information fields, enabling automated analysis, recurring queries, and integration with external programs or databases.
    \item[B16] Add an automated interface for synchronization with the EU AI database, including export and import functions to support data alignment and reduce duplicate reporting.
    \item[B17] Increase transparency on how feedback submissions are reviewed and followed up, including responsible institutions, review steps, timelines, and whether users receive a response or status update.
    \item[B18] Introduce regular user satisfaction surveys to assess whether the register is accessible, understandable, useful, and responsive to the needs of different user groups.
    \item[B19] Publish concise security and data protection documentation, including information on encrypted transmission, access control, identity and access management, and compliance with national and European data protection standards, without disclosing sensitive technical details.
\end{itemize}

\begin{table*}[t]
  \centering
  \setlength{\tabcolsep}{4pt}
  \renewcommand{\arraystretch}{2.2}
\begin{tabular}{P{7.8cm} m{3cm} >{\centering}m{1.4cm} @{\hspace{4pt}} | @{\hspace{4pt}} >{\centering}m{1.4cm} m{3cm}}
    \toprule
   % \cmidrule(lr){2-3} \cmidrule(lr){4-5}
    \textbf{Secondary Goals}
      & \multicolumn{2}{c}{\textbf{MaKI}}
      & \multicolumn{2}{c}{\textbf{PLS}} \\
    \midrule \midrule
  \textbf{S1: Promoting internal knowledge exchange and innovation within public authorities} & & & & \\
    \subgoal{S1a: Provision of basic information and contact details to enable further knowledge exchange}
      & \actionitem{MA1-4, B1-3}
      & \raisebox{-0.4\height}{\pie{0.6}{77}{3}{17}{3}}
      & \raisebox{-0.4\height}{\pie{0.6}{67}{0}{33}{0}}
      & \actionitem{LS1-7, B1-3} \\
    \subgoal{S1b: Establishment of a best-practice network among public authorities}
      & \actionitem{B4, B5, B20, B21}
      & \raisebox{-0.4\height}{\pie{0.6}{43}{14}{29}{14}}
      & \raisebox{-0.4\height}{\pie{0.6}{43}{14}{29}{14}}
      & \actionitem{B4, B5, B20, B21} \\
    \midrule
  \textbf{S2: Collaboration between public administration businesses and academia} & & & & \\
    \subgoal{S2a: Publication of AI systems used in companies and research institutions}
      & \actionitem{MA1-5, B1-4, B6}
      & \raisebox{-0.4\height}{\pie{0.6}{70}{3}{24}{3}}
      & \raisebox{-0.4\height}{\pie{0.6}{68}{0}{32}{0}}
      & \actionitem{LS1-7, B1-4, B6} \\
    \subgoal{S2b: Provision of basic information and contact details to enable further knowledge exchange}
      & \actionitem{MA1-4, B1-3}
      & \raisebox{-0.4\height}{\pie{0.6}{77}{3}{17}{3}}
      & \raisebox{-0.4\height}{\pie{0.6}{67}{0}{33}{0}}
      & \actionitem{LS1-7, B1-3} \\
    \subgoal{S2c: Exchange of results and innovations in open networks}
      & \actionitem{B4-6, B21}
      & \raisebox{-0.4\height}{\pie{0.6}{55}{9}{36}{3}}
      & \raisebox{-0.4\height}{\pie{0.6}{55}{9}{36}{3}}
      & \actionitem{B4-6, B21} \\
    \midrule
  \textbf{S3: Minimizing administrative bureaucratic and data-related effort} & & & & \\
    \subgoal{S3a: Introduction of standardized information fields for AI systems}
      & \actionitem{MA1-4, B1-3}
      & \raisebox{-0.4\height}{\pie{0.6}{78}{3}{16}{3}}
      & \raisebox{-0.4\height}{\pie{0.6}{65}{3}{32}{0}}
      & \actionitem{LS1-7, LS9, B1-3} \\
      \subgoal{S3b: Digitized automated and standardized processes for data collection verification and publication}
      & \actionitem{MA6, MA7, B7, B9-19}
      & \raisebox{-0.4\height}{\pie{0.6}{24}{12}{12}{52}}
      & \raisebox{-0.4\height}{\pie{0.6}{28}{24}{12}{36}}
      & \actionitem{LS8, LS9, B7, B9-19} \\
      \bottomrule
  \end{tabular}
  \caption{Secondary goals and subgoals of an AI register and how they are fulfilled by MaKI and PLS, along with action items needed for their fulfillment.}
  \label{tab:subgoals}
\end{table*}

\end{appendix}

\end{document}